\documentclass[namedreferences]{solarphysics}
\pdfoutput=1
\usepackage[hyperref,optionalrh]{spr-sola-addons} 
\usepackage{graphicx}        
\usepackage{amssymb}        
\usepackage{color}           
\usepackage{breakurl}        
\usepackage{natbib}
\usepackage{draftwatermark}

\newcommand{\evento}{{\tt SOL2012-03-13}}

\newcommand{\xwps}{WPS_\otimes(i,s)}

\SetWatermarkText{Draft}
\SetWatermarkScale{5}

\begin{document}

\begin{article}

\begin{opening}

\title{Analysis of intermittency in submillimeter radio and hard X-Rays during the impulsive phase of a solar flare}

\author{G.~\surname{Gim\'enez de Castro}$^{1,2}$\sep
        P.J.A~\surname{Sim\~oes}$^3$\sep
        J.-P.~\surname{Raulin}$^{1}$\sep
        O.M.~\surname{Guimar\~aes Junior}$^{1}$
      }
\runningauthor{Gim\'enez de Castro et al.}
\runningtitle{Intermittency in HXR and Submillimeter emission}

   \institute{$^{1}$ CRAAM, Universidade Presbiteriana Mackenzie, S\~ao Paulo, Brazil;
		email:~\url{guigue@craam.mackenzie.br};
              $^{2}$ IAFE, Universidad de Buenos Aires, Buenos Aires, Argentina;
              $^{3}$ SUPA School of Physics and Astronomy, University of Glasgow, Glasgow G12 8QQ, UK
            }

\begin{abstract}
We present an analysis of intermittent processes occurred during the
impulsive phase of the flare \evento, using hard X-rays and
submillimeter radio data. Intermittency is a key characteristic in
turbulent plasmas and have been analyzed recently for hard X-ray data
only. Since in a typical flare the same accelerated electron
population is believed to produce both hard X-rays and
gyrosynchrotron, we compare both time profiles searching for
intermittency signatures. For that we define a cross-wavelet power
spectrum, that is used to obtain the local intermittency measure or
$LIM$. When greater than three, the square $LIM$ coefficients indicate
a local intermittent process.  The $LIM^2$ coefficient distribution in
time and scale helps to identify {\em avalanche} or {\em cascade}
energy release processes. We find two different and well separated
intermittent behaviors in the submillimeter data: for scales greater
than 20~s, a broad distribution during the rising and maximum phases
of the emission seems to favor a {\em cascade} process; for scales
below 1~s, short pulses centered on the peak time, are representative
of {\em avalanches}. When applying the same analysis to hard X-rays,
we find only the scales above 10~s producing a distribution related to
a {\em cascade} energy fragmentation. Our results suggest that
different acceleration mechanisms are responsible for tens of keV and
MeV energy ranges of electrons.

\end{abstract}
\end{opening}


\section{Introduction}

Hard X-rays (HXRs) and the radio continuum $> 5$~GHz during the
impulsive phase of solar flares are believed to be produced by
accelerated electrons.  HXR emission is non-thermal bremsstrahlung
from accelerated electrons with energies above 10~keV
\citep{Brown:1971,Kontaretal:2011}. The same electrons spiraling along
the magnetic field lines produce the broad-band gyrosynchrotron (GS)
continuum above a few GHz.  On the other hand, the relativistic
electrons produce radio emission above 30~GHz, which is, in most
flares, optically thin
\citep{WhiteKundu:1992,Ramatyetal:1994,Trottetetal:2015}.

Many works have been devoted to the comparison between HXR and radio
observations during solar flares \cite[\textit{e.g.}][and references
  therein]{Whiteetal:2011}. In this work we focus on one aspect of the
impulsive phase, namely, its intermittent characteristic. And for that
we choose a particular event: the flare \evento, which was already
analyzed by \inlinecite{Kaufmannetal:2013} and
\inlinecite{Trottetetal:2015}. Our choice is related to the fact that
HXR data $\le 300$~keV show an impulsiveness which is not observed at
high radio frequencies, although the latter present fast sub-second
pulses. We apply in this work the wavelet formalism to describe the
intermittent processes as it was first used by
\inlinecite{Farge:1992}, who defined the local intermittency measure
($LIM$), and later on used by \inlinecite{Brunoetal:1999} for the
analysis of spatial characteristics of turbulent plasmas. More
recently \cite{ConsoliniChang:2002} used the same formalism to find
temporal intermittencies in the terrestrial magnetosphere, and
\inlinecite{DinkelakerMacKinnon2013A} analyzed the HXR flux produced
during the impulsive phase of flares.  The latter applied the $LIM$
analysis to a series of flare emission models produced by an energy
fragmentation, or cascade, and by an avalanche process, obtaining the
main characteristics expected from these two different turbulent
sources.  The modeling results were applied to sub-second time
resolution HXR data to identify cascade and avalanche mechanisms
\citep{DinkelakerMacKinnon2013B}.

We use here radio data from the \textit{Solar Submillimeter Telescope}
\cite[SST,][]{Kaufmannetal:2008} at 212~GHz and data from the \textit{Fermi
Gamma-ray Burst Monitor} \cite[GBM,][]{Meeganetal:2009}, both with
sub-second high time resolution in order to compare the
occurrence and characteristics of the intermittent processes. For this
purpose we redefine the $LIM$ representation using a cross-wavelet
power spectrum, and submit the data to a $3\sigma$ significance level
test. The formalism and method are presented in Section
\ref{sec:formalism}, an overview of the flare is described in Section
\ref{sec:overview}, while the analysis is presented in Section
\ref{sec:lim}, with a discussion and final remarks in Section
\ref{sec:discussion}.

\section{Intermittency and Wavelet Transform}
\label{sec:formalism}

Wavelet transform is both localized in time and in frequency (or time
scale). This inherent capacity makes it ideal for identifying
intermittent processes which are random in time and variable in
frequency \citep{Farge:1992}. \inlinecite[][hereafter
  TC98]{TorrenceCompo:1998} wrote an excellent and very comprehensive
review on wavelets offering also a set of ready-to-use programs for
computing the continuous wavelet transform.  In geophysics and
astronomy these tools became standards that can be found in many
different computing languages like Python, Matlab, and IDL, among
others. It should be noted, however, that before Torrence and Compo,
\cite{Farge:1992} recommended the use of continuous (non-orthogonal
and redundant) wavelets for the analysis of turbulent fluids. In
Farge's words \citep{Farge:1992} [...]  {\em the continuous wavelet
  transform is better suited because its redundancy allows good
  legibility of the signal's information content}. Therefore we use
TC98 programs to compute the continuous wavelet transform of our data
sets.

\begin{figure}
\includegraphics[width=\textwidth]{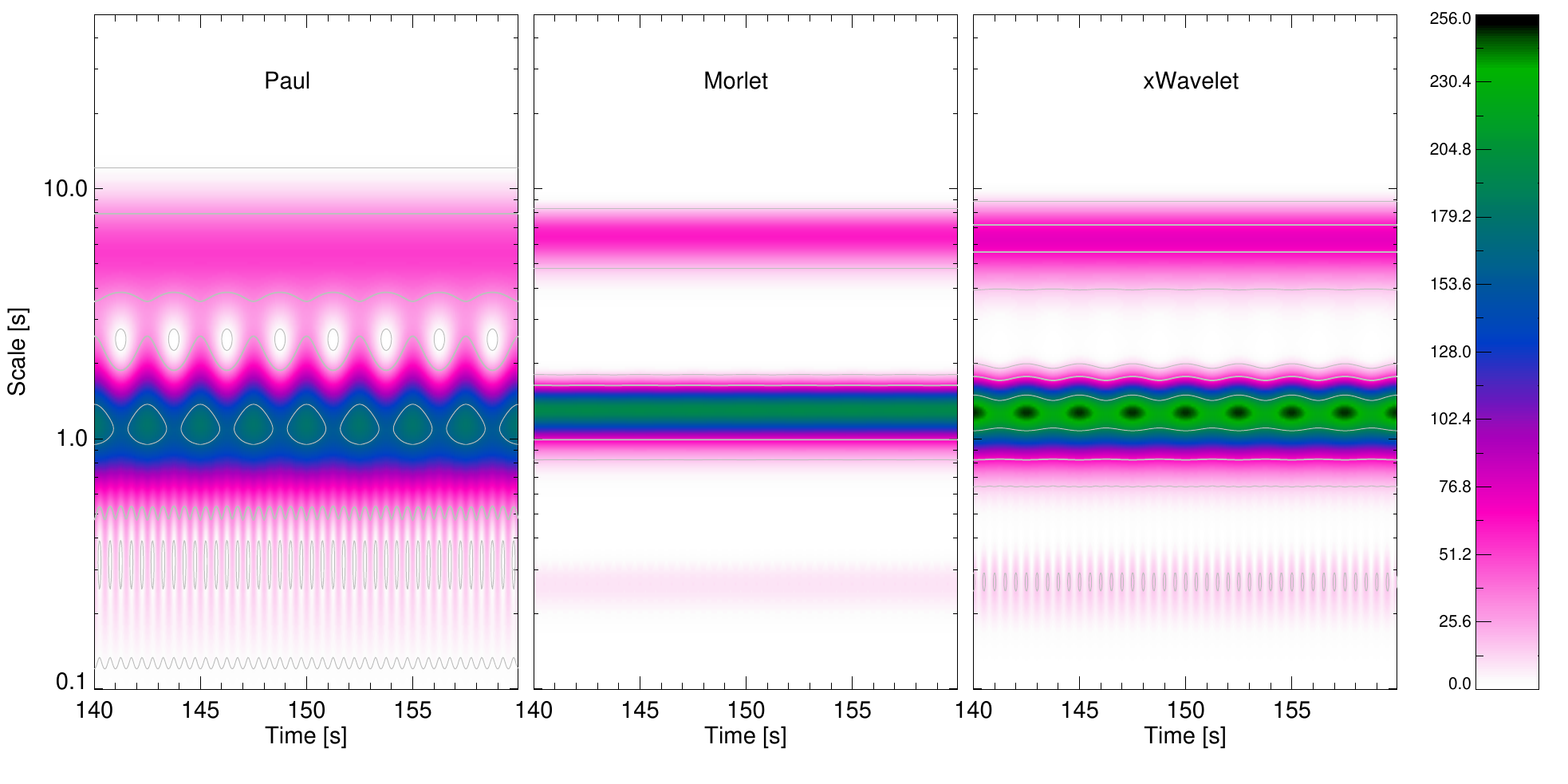}
\caption{Left: a color coded representation of the wavelet power
  spectrum (WPS) in function of time and scale for the Paul
  WF. Middle: the same for the Morlet WF. Right: the same for the
  cross wavelet transform $\xwps$ (see text). On the extreme right the
  color code is shown.}
\label{fig:xWaveletComparison}
\end{figure}

By definition, given a
discrete time series $x_i = x(t_i) \ , i=\{0\ldots N-1\}$, where $t_i$
represents the time bins, the wavelet coefficient 
$W(i,s)$  for the time scale $s$ is defined  as
 \begin{equation}
W(i,s) = \sum_{j=0}^{N-1}  x_j \psi^* \left [ \frac{(j-i)\delta t}{s}\right ] \ ,
\label{eq:CWT}
\end{equation}
where $\psi^*$ is the wavelet function conjugate, and $\delta
t=t_{i+1}-t_i$, is the time step of the series, assumed constant for
the analyzed period (TC98). There are different wavelet functions
(WF), which can have different impact on the data analysis. All of
them are governed by the uncertainty principle
$$\frac{\Delta t}{\Delta s} > \mbox{constant} \ , $$ 
($\Delta$ means uncertainty), that precludes increasing the
precision over the time and frequency domains simultaneously.
However, some WF have a better precision in one domain than in the
other. TC98 programs provide the complex Morlet and Paul WF, and also the
real-valued {\em derivative of a Gaussian} (DOG). Morlet WF is
better localized in the frequency domain while the Paul WF is more
precise in time.  As an example we produced a time series represented
by the following parametric formula
\begin{equation}
x_i(t_i) = A \sin(2\pi t_i/T_1) + B \sin(2\pi t_i/T_2) + C\sin(2\pi t_i/T_3) \ , \nonumber
\end{equation}
where $A$, $B$ and $C$ are constants, $T_1 = 0.4$~s, $T_2=2.0$~s and
$T_3=10.0$~s. The time bins are formed by the sequence $t_i =
0.04i$~s. We computed the wavelet power spectrum
\footnote{This is the {\em local} power spectrum. The {\em global}
  power spectrum, not used in this work, is the time averaged of the
  local wavelet power spectrum for every scale $\langle
  WPS(s)\rangle_t$.}
$$WPS(i,s) = |W(i,s) \ W^*(i,s)| \ , $$ using Paul and Morlet WFs, and
we call them $WPS_P$ and $WPS_M$ respectively. In figure
\ref{fig:xWaveletComparison}, the left panel shows the color coded
$WPS_P$.  As expected, Paul WF identifies the fast pulses with 0.4 s
and 2.0~s periodicity.  It also recognizes the existence of a periodic
pulsation with a time scale around 6.0~s. Although the input signal is
discrete in frequency, the figure seems to support the idea of a
continuous variation in scale from around 10~s to 0.1~s. This artifact
is created by the relatively coarse frequency resolution of the Paul
wavelet.

The middle panel of Figure \ref{fig:xWaveletComparison} shows the
$WPS_M$. Contrary to the previous case, we observe three discrete
horizontal bands, corresponding to the three time scales in the input
signal $x_i$, but there is no identification of the occurrence time of
individual pulses.  This is the expected result for a Morlet wavelet,
working fine in the scale domain and coarse in the time domain.
Finally, in the right panel we present the color coded cross-wavelet
power spectrum defined as
\begin{equation}
\xwps = | W_P(i,s) \ W^*_M(i,s)| \ ,
\end{equation}
In TC98, $\xwps$ is defined for two different time series, using the
same WF. We find that the same principle can be used for two different
wavelet representations of the same input data. We note that this
correlation can be carried on because both wavelet functions use the
same time and scale bins for the convolution with the signal.  In
Figure \ref{fig:xWaveletComparison} it can be observed that the
$\xwps$ has the capacity of Paul WF to identify individual peaks while
still being able to separate the different time scales, a
characteristic of the Morlet WF.  The caveat of the $\xwps$ is that we
cannot attribute a Fourier frequency to a time scale, since Fourier
Frequencies depend on the WF (TC98). Time scales, on the contrary, are
independent of the WF.

\inlinecite{Farge:1992} introduced the concept of $LIM$ analyzing the
spatial structure of turbulent fluids.  \inlinecite[][hereafter
  DMac13]{DinkelakerMacKinnon2013A,DinkelakerMacKinnon2013B} applied
$LIM$ to the study of HXR flare data, with the aim of
identifying episodes of energy release in the time series. $LIM$ is
obtained by DMac13 from the wavelet power spectrum
\begin{equation}
LIM(i,s) = \frac{WPS(i,s)}{\langle WPS(i,s)\rangle_t} \ ,
\label{eq:LIM}
\end{equation}
where $\langle\rangle_t$ is the mean over the time interval for a
fixed scale $s$. Intermittency reflects the fact that energy is not
evenly distributed \citep{Farge:1992}, \textit{i.e.} whenever
$LIM(i,s) >1$.  \inlinecite{Brunoetal:1999} clarifies this idea,
stating that $LIM(i,s) > 1$ means that this fluctuation has more
energy than expected for a normal distribution and therefore indicates
a departure from a Gaussian distribution of fluctuations. When the
departure happens, neither the mean, nor the standard deviation of the
distribution represent well the sample, and higher order moments are
needed, like the kurtosis $K_4$. A Gaussian random variable has
$K_4=3$, larger values denote a flatter distribution whose wings are
over-represented respect to the central
mean. \inlinecite{Meneveau:1991} states that $LIM^2$ is an estimate
for $K_4$, therefore the condition of an intermittent behavior at time
$t_i$ and scale $s$ are given by
$$LIM(i,s)^2 > 3 \ . $$

DMac13 use the $LIM$ analysis to also characterize the energy release
fragmentation process. There are two main pictures: the {\em cascade}
or top-down, and the {\em avalanche} or bottom-up. In the first case,
the energy is released and fragmented during the time progression. On
the other hand, an avalanche is related to the self-organized critical
(SOC) state and first applied to solar flares by
\inlinecite{LuHamilton:1991} (for a didactic review see
\opencite{Charbonneauetal:2001}). In this picture, small events
trigger other small events, resulting in an energy avalanche. There
are no privileged theory and probably both processes take place during
flares. But evidences for one or the other at different times may
bring clues about the origin of the energy. Based on time series
simulations, DMac13 conclude that \textit{stalactite} structures
observed in a $LIM(i,s)^2$ graph, \textit{i.e.} vertical patches where the top
is larger than the bottom, may indicate cascade processes, while {\em
  stalagmites}, \textit{i.e.} patches with a larger bottom and a narrower top,
are representative of avalanches.

\subsection{The cross-wavelet {\em LIM}}
\label{sec:xwavelet}

In this work we define the cross-wavelet $LIM$ as
\begin{equation}
LIM_\otimes(i,s) = \frac{|WPS_\otimes(i,s)|}{\langle |WPS_\otimes(i,s)|\rangle_t} \ . 
\end{equation}
The choice of a WF involves also the choice of its characteristic
parameters. The Morlet WF is a plane wave modulated by an exponential.
Its main parameter is the central wave frequency $\omega_\circ$. After
some tests we found that $\omega_\circ=4$ produced the best results in
terms of frequency localization for our data sets.  On the other hand,
the Paul WF is a complex rational polynomial of order $m+1$.  Typical
values for $m$ are between 2 and 4; we choose $m=3$ for all our
calculations.

One crucial step to recognize intermittent structures is a
significance test. We compute the 99.7\% significance level $SL_X(s)$
(\textit{i.e.} there are 0.3\% chances that the result is purely random) for
every scale $s$, where $X$ can be $P$ (Paul) or $M$ (Morlet). The
significance level depends on the standard deviation $\sigma$ of data
and on a noise spectrum model (TC98).  For the present analysis we
adopt a red-noise defined by TC98 as
\begin{equation}
P_k = \frac{1-\alpha^2}{1+\alpha^2-2\alpha \cos(2\pi k /N)} \ ,\nonumber
\end{equation}
where $N$ is the number of time bins and $k = {1,\ldots N/2}$ is the
frequency index. The parameter $\alpha$ regulates the spectrum:
$\alpha=0$ gives a white-noise. Here we use, as suggested by TC98
\begin{equation}
\alpha = \frac{R_1 + \sqrt{R_2}}{2} \ , \nonumber
\end{equation}
where $R_1$ and $R_2$ are the auto-correlation coefficient with a lag
equal to 1 and 2 respectively. Since we are using a cross-wavelet we
have to compare the cross-wavelet coefficients with the product of the
$SL_X$
\begin{equation}
SL_\otimes(s) = \sqrt{SL_P(s) \ SL_M(s)} \ . \nonumber
\end{equation}
Therefore a local cross-wavelet power is not random if and only if
(TC98)
\begin{equation}
WPS_\otimes(i,s) > SL_\otimes(s) \ . \nonumber
\end{equation}

In summary the steps to compute the $LIM_\otimes(i,s)$ are
\begin{enumerate}
\item Obtain the standard deviation $\sigma$. For this purpose we
  select a data subset outside the flare time interval. The reasoning
  is that using the flaring period, we will introduce in the
  computation real features.
\item Compute the 99.7\% cross-significance level $SL_\otimes(s)$.
\item Compute the Paul $W_P(i,s)$ and Morlet $W_M(i,s)$ cross-wavelet
  decompositions of the flare data.
\item Obtain the cross-wavelet spectrum $WPS_\otimes(i,s)$ and the local intermittency  
      measure $LIM_\otimes(i,s)$.
\item For a given $i_\circ$ and $s_\circ$ verify that
      \begin{enumerate}
      \item $WPS_\otimes(i_\circ,s_\circ) \ge SL_\otimes(s_\circ)$ (significance test) and
      \item $LIM_\otimes(i_\circ,s_\circ)^2 \ge 3$ (non-Gaussianity test).
      \end{enumerate}
      If these two conditions are not simultaneously fulfilled, then
      attribute 0 to $LIM_\otimes(i_\circ,s_\circ)$. Repeat this
      process for all $(i,s)$.
\end{enumerate}
Following the above steps we ensure that the resulting
$LIM_\otimes(i,s)^2 \ne 0$ are above the significance level and
correspond to an intermittent process. 

\subsection{Tests}
\newcommand{\xLIMsq}{LIM_\otimes^2}

\begin{figure}
\begin{center}
\includegraphics[width=\textwidth]{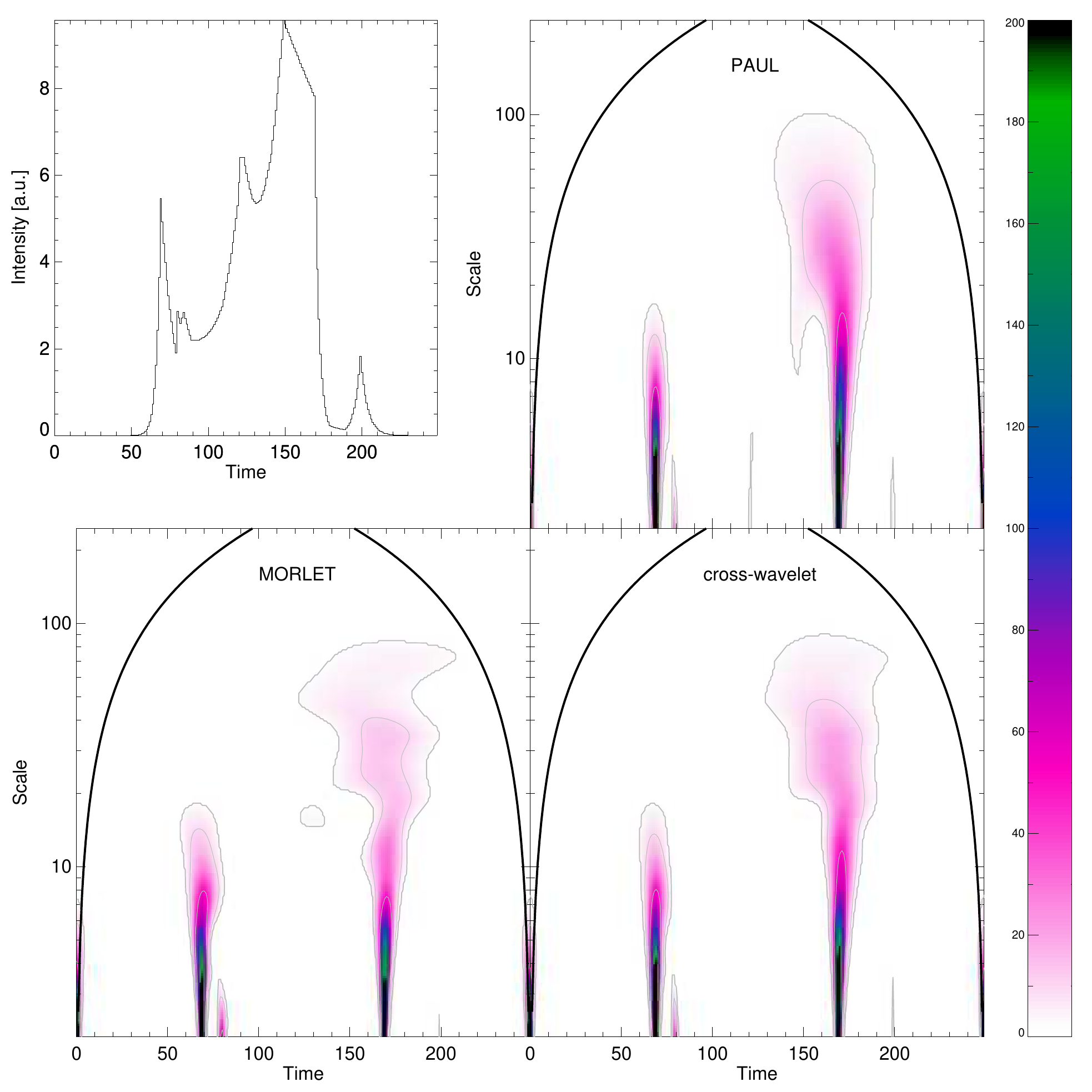}
\end{center}
\caption{$LIM^2$ representation for simulated cascade data. In the top-left panel
the light curve is shown. The other three panels present the $LIM^2$
analysis for Paul, Morlet and cross-wavelet representations.}
\label{fig:CantorSet}
\end{figure}

 Following DMac13, we applied the $\xLIMsq$ method on simulated
  data for the two scenarios: {\em i)} cascades and {\em ii)}
  avalanches. A cascade process is simulated by means of a Cantor set
  defined by the iterative formula
  \citep{Argouletal:1989,DinkelakerMacKinnon2013A}
$$f(t_0) = f_\circ \quad , \ f(t_i) = \alpha f(t_{i-1}) = \alpha
  f_\circ^\alpha\quad , \ i=\{1, \ldots N\} \ . $$ 
In our approach $\alpha=\alpha(t_i)$, a function of the time step
$t_i$ and can be either positive or negative.  Figure
\ref{fig:CantorSet} shows on the top-left panel the simulated light
curve. The three other panels are the $LIM^2$ representations for
Paul, Morlet and the cross-wavelet decompositions (see labels in each
panel). Paul and Morlet show intermittencies at time=50, 80, 170 and
200 s (the latter very weak in Morlet) and consequently, the $\xLIMsq$
shows these four intermittency structures.  In the Paulo $LIM^2$
representation, there is also a weak structure for time=120~s not seen
in Morlet, that corresponds to a peak in the light curve at the same
time. Moreover, there is a short {\em stalactite} with peak at
time=140~s and with scale minimum 10~s, which is not visible in
Morlet. On the contrary, Morlet shows a circular patch for scale
$\sim$ 15~s around the same time.  None of these structures seem to be
related to the light curve. And the $\xLIMsq$ does not show any of
them.  In overall we see that the $\xLIMsq$ agrees with DMac13 that
cascades produce {\em stalactites} but it gives a cleaner
representation than a pure Paul decomposition.

\begin{figure}
\begin{center}
\includegraphics[width=\textwidth]{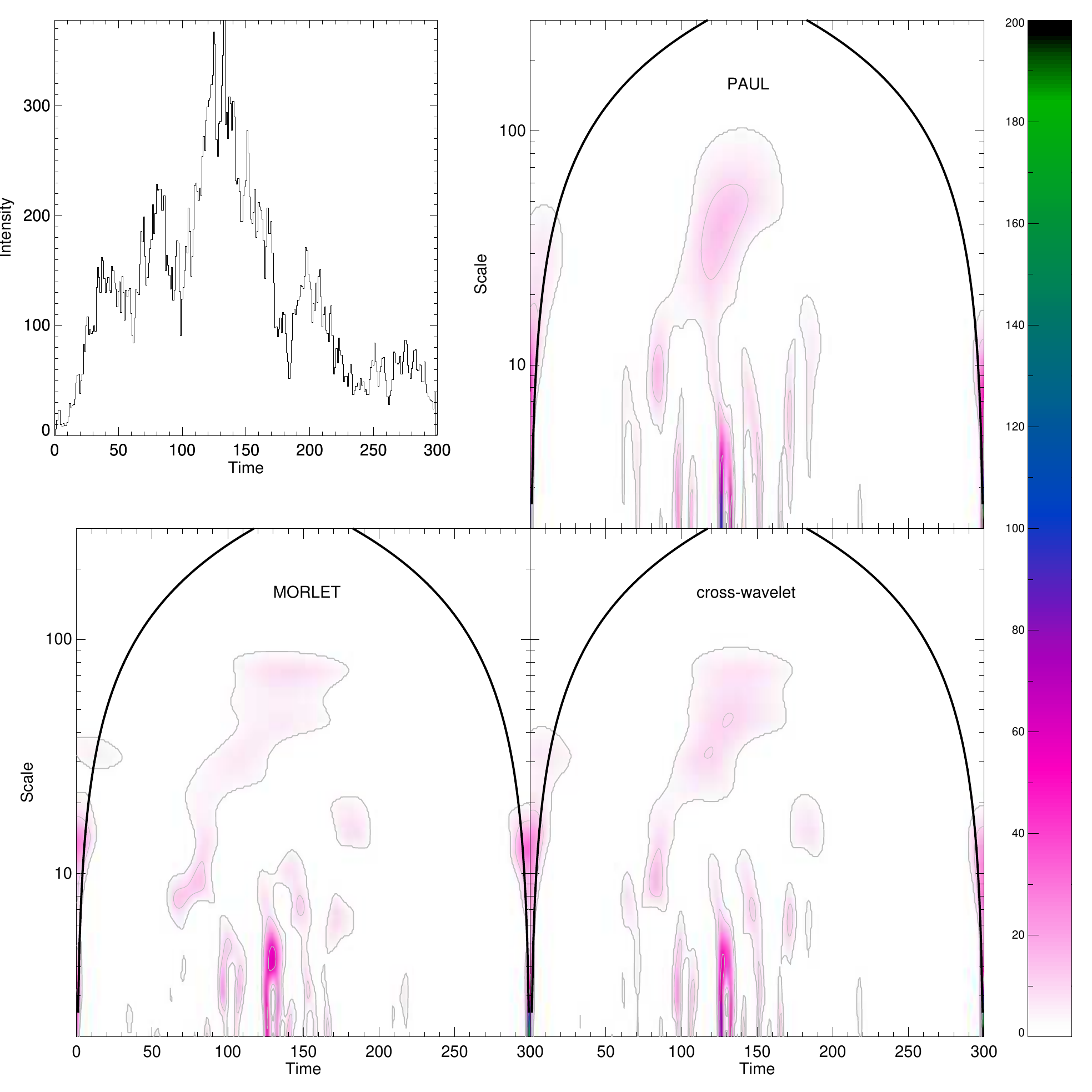}
\end{center}
\caption{The same as in Figure \ref{fig:CantorSet}, but for a simulated avalanche
light curve.}
\label{fig:Avalanches}
\end{figure}

Simulated avalanche data, produced by a Markovian cellular automata
and used in DMac13 for tests, were kindly provided by Dr. Alexander
MacKinnon. In Figure  \ref{fig:Avalanches}, top-left panel, we show the
test data (the same of Fig. 4(c) in
\opencite{DinkelakerMacKinnon2013A}).  As in the previous case, the
three different representations of the $LIM^2$ analysis have overall
the same characteristics. However, the Paul decomposition shows a link
between scales from 100~s to 1~s (see around time=120~s), creating a
hierarchy of scales that Morlet does not show because the intermediate
scales between 10 and 30~s are missing.  The $\xLIMsq$ uses Morlet
information and removes the false link between two different scale
hierarchies.

We emphasize that the global characteristics of cascade and avalanche
processes obtained with the cross-wavelet approach are similar to
those obtained by DMac13 using the Paul wavelet.

\section{Overview of the flare}
\label{sec:overview}

The flare \evento\ occurred in the active region AR 11429 located at
N18W50 (00:00 UT). It was classified as M7.9 in soft X-rays (SXRs) and
1B in H$\alpha$. For a complete description of the observational data see
\inlinecite{Kaufmannetal:2013} and \inlinecite{Trottetetal:2015}; both
works focus on the analysis of the 30 THz emission. In particular
\inlinecite{Trottetetal:2015} concluded that its origin is compatible
with thermal radiation that originate from two different sources: 80\% of
the emission comes from a $T \sim 8000$~K optically thin source at
chromospheric heights, while the remaining 20\% is emitted by an
optically thick photospheric source with $T\sim 6000$~K.  Furthermore,
\inlinecite{Trottetetal:2015} interpret the chromospheric temperature
excess that produces the 30 THz emission with the energy deposited by
non-thermal electrons. The same accelerated electron distribution is
responsible for the thick-target bremsstrahlung hard X-ray and the GS
submillimeter radiation. For the HXR emission
\inlinecite{Trottetetal:2015} found an electron distribution with two
spectral indices, and a break energy at around 400~keV. On the other
hand, the GS optically thin emission is compatible with a source of
electrons with energies $800 \le E_e \le 10^4$~keV, having a trapping
time of around 2~s, gyrating in a magnetic field with effective
intensity $B_\circ \sim 700$~G. All these GS parameters were deduced
for an homogeneous emitting source with a presupposed size.

\begin{figure}
\includegraphics[width=\textwidth]{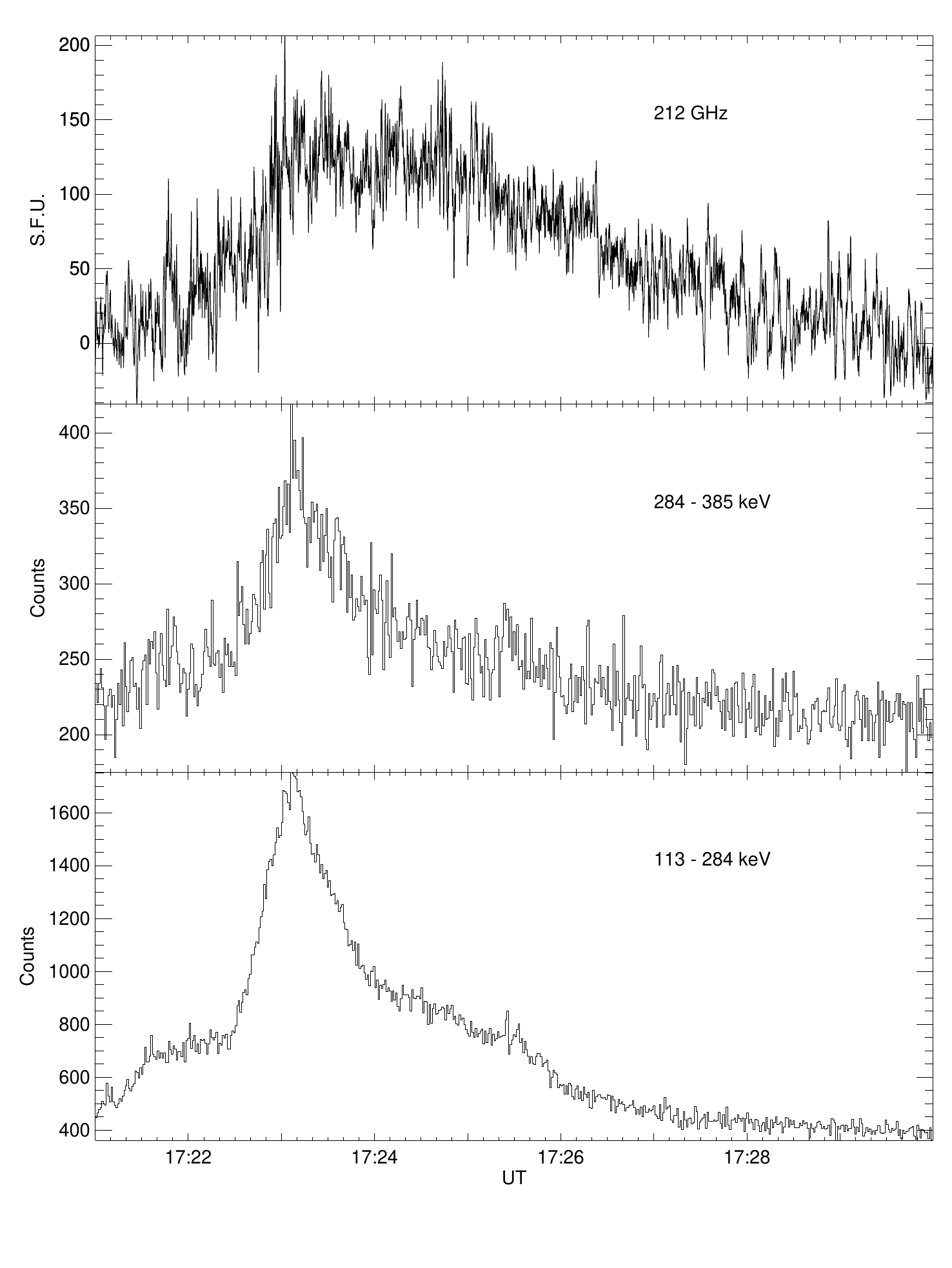}
\caption{Top: flux density at 212 GHz with 40 ms time
  resolution. Middle and bottom: \textit{Fermi} HXR counts at the
  indicated energy bands with 1024~ms time resolution.}
\label{fig:TimeProfiles}
\end{figure}

In this work we are interested in the comparison between HXR and
submillimeter emission.  We use the obtained flux at
212~GHz\footnote{By definition, emission at 212 GHz ($\lambda=1.4$~mm)
is not submillimeter. However, here, as in other publications, we
consider it submillimeter because its origin are $>$~MeV electrons,
the same that produce the $\lambda < 1$~mm emission. See
\textit{e.g.} \inlinecite{Ramatyetal:1994,Trottetetal:2015}.} with 40~ms time
resolution.  We compare it with the \textit{Fermi} HXR counts for the bands
$113.25 \le E_{low} \le 294.65$~keV and $ 284.65 \le E_{high} \le
385.26$~keV with 1024~ms time resolution. These are the highest energy
bands for which a good signal-to-noise ratio is found.  In Figure
\ref{fig:TimeProfiles} we show the flux density at 212 GHz (top panel)
and the HXR counts (middle and bottom panels).  The submillimeter time
profile is less impulsive than the HXR. Indeed, the 212 GHz peak time
can be hardly identified because it has a {\em plateau} between
17:23:00 until 17:24:30~UT, but nevertheless, the GS spectrum is
compatible with HXR \citep{Trottetetal:2015}.

\section{Local Intermittency Measure}
\label{sec:lim}

\begin{figure}
\includegraphics[width=\textwidth]{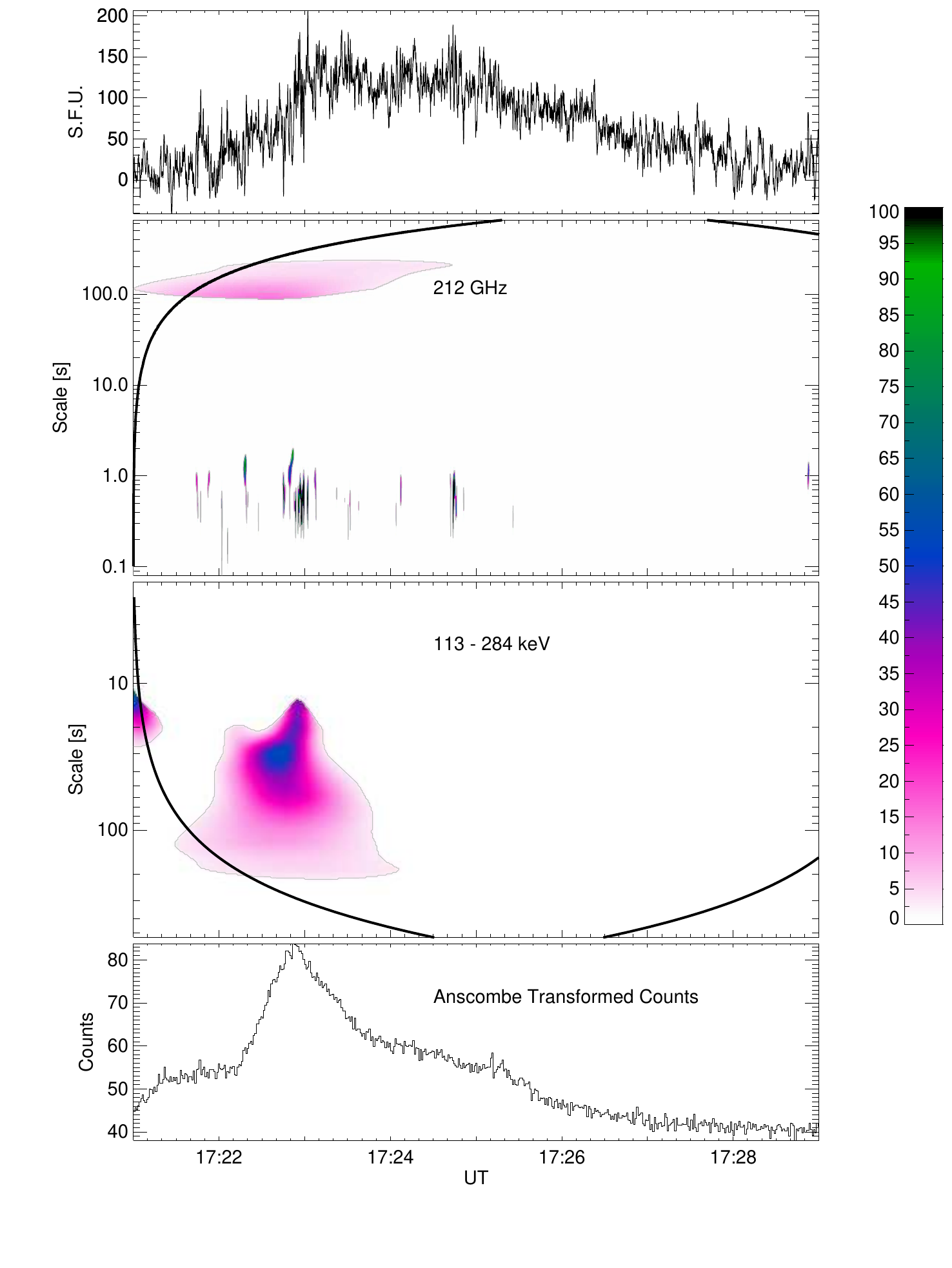}
\caption{From top to bottom: Flux density at 212~GHz; color coded
  $\xLIMsq$ for submillimeter data; color coded $\xLIMsq$ for the
  $E_{low}$ \textit{Fermi} band; and counts for $E_{low}$ energy band.  HXR data 
  are converted to a Gaussian distribution using the Anscombe transform.
  Note that only values greater than three and which correspond to WPS
  coefficients above the significance level are colored and that the
  ordinate axis of the \textit{Fermi} data is inverted. The thick black curves
  in the $\xLIMsq$ represent the {\em cone of influence}, beyond which,
  the border effects make the wavelet transform not representative of
  the data. On the right, the color code for the $\xLIMsq$
  representations. }
\label{fig:LIM-Fermi_low}
\end{figure}

Figures \ref{fig:LIM-Fermi_low} and \ref{fig:LIM-Fermi_high} present
the $\xLIMsq$ for the submillimeter and HXR time profiles. Since HXR
data have a Poisson statistics we used the Anscombe transformation
\citep[see \textit{e.g.} ][]{MurtaghStarck:2006p42}
\begin{equation}
\label{eq:anscombe}
{\cal C}_i = 2 \sqrt{c_i + \frac{3}{8}} \ ,
\end{equation}
to convert the original counts $c_i$ to ${\cal C}_i$ with Gaussian
distribution and variance $\sigma^2=1$. In this way we can use the
TC98 formalism which was deduced for Gaussian distributions only. We
note that only significant wavelet coefficients and $\xLIMsq \ge 3$
are shown in accordance with the method described in
Section \ref{sec:xwavelet}.  We also note that the HXR $\xLIMsq$ are
represented with the ordinates axis inverted.

\subsection{Submillimeter $\xLIMsq$}

We identify two different intermittent components in the submillimeter
data.  Between 17:21:00 and 17:24:30~UT, coincident with the signal
rising and maximum, a weak intermittency ($\xLIMsq \sim 3$) is
concentrated in scales between 100~s and 200~s.  Besides this broad
patch structure, a sequence of short non-continuous intermittencies
with scales $\le 2$~s is observed, with the following characteristics:
\begin{enumerate}
      \item They are detached from the broad patch,
      \item the most intense $\xLIMsq$ coefficients correspond to the
        sub--second scales, 
      \item they are not uniformly distributed, but centered around
        the  HXR  at 17:22:50 UT.
      \item More important: the rate and amplitude (measured by
        $\xLIMsq$) of intermittencies increases with the intensity of
        the event, reaching values for $\xLIMsq=380$ at the scale
        0.7~s.
\end{enumerate}
Following DMac13, the whole group presents the characteristics of an
avalanche process for the shortest scales.

\subsection{HXR $\xLIMsq$}

\begin{figure}
\includegraphics[width=\textwidth]{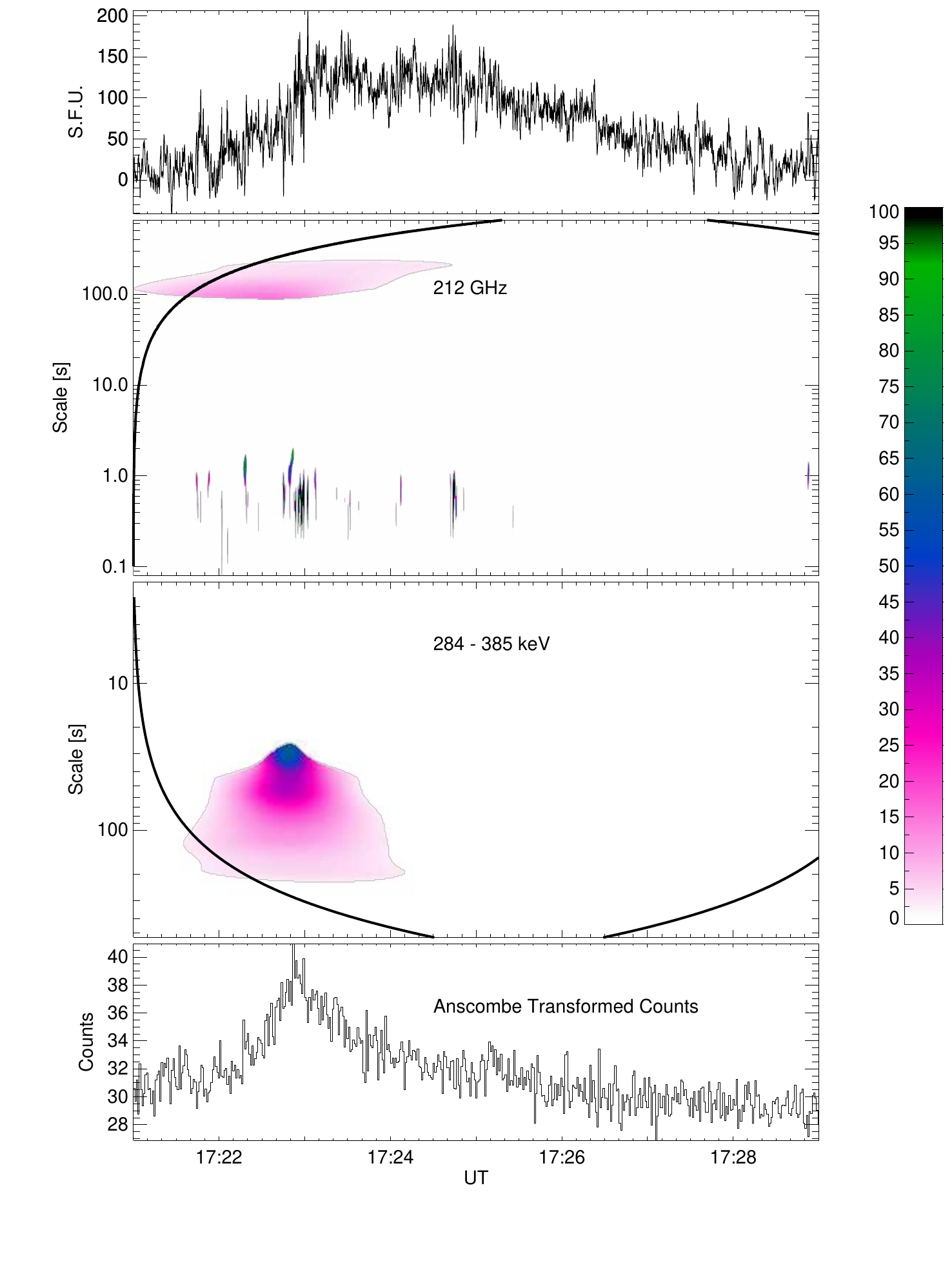}
\caption{Same as the Figure \ref{fig:LIM-Fermi_low} for the
  $E_{high}$ energy band.}
\label{fig:LIM-Fermi_high}
\end{figure}

The bottom panel of Figure \ref{fig:LIM-Fermi_low} presents the
$\xLIMsq$ coefficient for the HXR $E_{low}$ energy band.  We have
inverted the scale axis to ease the comparison with the submillimeter,
therefore, in this situation, {\em stalactites} are bottom-up
structures, meanwhile {\em stalagmites} are the opposite
top-to-down. A broad patch, coincident with that of the submillimeter
emission, is observed, between the scales 200~s, in the edges, down to
10~s at peak time (17:22:50 UT).  The maximum $\xLIMsq$ value for this
time period is 44 at the scale 30~s. Figure \ref{fig:LIM-Fermi_high}
shows in its bottom panel the HXR $E_{high}$ energy data $\xLIMsq$
representation. In this case, a similar broad patch is observed at
scales between 200~s in the edges and 20~s at peak time, with a
maximum of 62 at scale 28~s. These {\em stalactites} structures
suggest that the energy release went through a cascade process.

\section{Discussion and final remarks}
\label{sec:discussion}

We have developed a $\xLIMsq$ methodology based on a cross-wavelet
transform that simultaneously enhances the localization of
intermittencies in time and scale. We also use a significance test to
validate our results. Furthermore, in the $\xLIMsq$ matrix we only save
coefficients greater than three, to facilitate the intermittent processes
identification. We apply this methodology to compare HXR and
submillimeter emission during the \evento\ flare. Our analysis
revealed two different characteristics in the submillimeter
intermittency process:
\begin{enumerate}
\item A broad patch for scales between 100~s and 200~s  during the 
      rising and maximum phases of the event, and 
\item a sequence of disconnected intermittencies for scales below 2~s
  down to 0.1--0.2~s.
\end{enumerate}
On the contrary, the HXR $\xLIMsq$ appears mostly as a single
hierarchical structure, or broad patch, with scales from 200~s down to
10~s time coincident with the submillimeter broad patch.  

\inlinecite{Trottetetal:2015} used the electron flux $F_X(E)$ derived
from HXR spectra during the time interval 17:22:10 to 17:23:30 UT to
fit the GS mean spectrum. The obtained $F_X(E)$ is a double power law
with an energy break around 400~keV. By fitting the optically thin GS
spectrum, they conclude that it is produced by electrons with energies
above 800~keV. Unfortunately, \textit{Fermi} data above 300~keV are not
significant enough to be analyzed with the $\xLIMsq$
decomposition. Therefore we consider that both HXR energy bands used
in this work, are representative of the low-energy electrons while the
submillimeter data represent the high energy electrons. Thus, we can
conclude that the $\xLIMsq$ analysis shows two different energy
release regimes for the low and the high energies.

During the rising phase high energies are released with the cascade
mechanism producing emission scales from 200~s down to 100~s. Along
with the cascade there are episodes of energy release similar to
avalanches, with characteristic scales from 2~s down to 0.1~s,
that increase in number and intensity during the maximum of the
emission. At low energies, on the contrary, there are no clear
evidence of avalanches, and the cascade energy release produces scales
between 200~s and 10~s. We emphasize the fact that the $\xLIMsq$
clearly identifies the maximum of the submillimeter emission (which is
not evident from the time profile as we have said) and that it
coincide with the maximum in HXRs. This means that both energy release
regimes are associated by the trigger process. Moreover, $\xLIMsq$
submillimeter coefficients are much higher than their HXR
counterpart. Since the $\xLIMsq$ coefficients are obtained after a
normalization by the averaged wavelets, they can be compared even if
they correspond to different physical magnitudes.

From reasonability arguments, \inlinecite{Trottetetal:2015} concluded
that the electrons should be trapped inside the magnetic loop with a
characteristic time $\tau_{trap}=2$~s.  Trapping produces a delay in
the bremsstrahlung time profile respect to the injection time
profile. On the other hand, the GS time profiles becomes enlarged
because of the trapping. This could be an explanation of larger scales
observed in the submillimeter intermittencies, however it would
require $\tau_{trap}$ of the order of tens of seconds, that imply a
lower magnetic field strength and a higher density of accelerated
electrons, beyond any {\em reasonable} scenario
\citep{Trottetetal:2015}. Therefore, if transport processes cannot
explain the difference in shape, a reason may be found in the
acceleration mechanism for the high energy particles.

In synthesis, we have shown in this work the existence of two
different intermittency components in the submillimeter emission of
the flare \evento\ originated in $>$~Mev accelerated electrons, and
that these two components characterize different intermittent
processes. The fast structures ($< 1$~s) occur around the emission
peak and show the more intense $\xLIMsq$ coefficients. On the other
extreme the slow components ($\ge 100$~s) are weak and coincide in
time with a similar structure observed in HXRs.  In a global view, we
find that a cascade process produces emission from 200~s to 10~s, from
Mev to hundred of keV.  We also find that avalanche processes are only
associated with Mev electrons. Our results suggest that different
acceleration mechanisms are responsible for keV and MeV energy ranges
of electrons.

\begin{acks}
G.~Gim\'enez de Castro and J.-P.~Raulin
acknowledge CNPq (contracts 300849/2013-3 and 312788/2013-4). This
research was partially supported by Brazil agency FAPESP (contract
2013/24155-3). Special thanks to Dr. Alexander MacKinnon, of University of
Glasgow, who kindly provided test data for our analysis. 
\end{acks}


\bibliographystyle{spr-mp-sola}
\tracingmacros=2
\bibliography{referencias}

 \IfFileExists{\jobname.bbl}{} {\typeout{}
\typeout{****************************************************}
\typeout{****************************************************}
\typeout{** Please run "bibtex \jobname" to obtain} \typeout{**
the bibliography and then re-run LaTeX} \typeout{** twice to fix
the references !}
\typeout{****************************************************}
\typeout{****************************************************}
\typeout{}}

\end{article} 

\end{document}